# Large-Scale Analysis of Pop-Up Scam on Typosquatting URLs

Tobias Dam

Institute of IT Security Research
St. Pölten University of Applied Sciences
Austria

tobias.dam@fhstp.ac.at

Damjan Buhov

Josef Ressel Center TARGET
St. Pölten University of Applied Sciences
Austria

damjan.buhov@fhstp.ac.at

Lukas Daniel Klausner

Institute of IT Security Research
St. Pölten University of Applied Sciences
Austria

mail@l17r.eu

Sebastian Schrittwieser

Josef Ressel Center TARGET
St. Pölten University of Applied Sciences
Austria

sebastian.schrittwieser@fhstp.ac.at

## ABSTRACT

Today, many different types of scams can be found on the internet. Online criminals are always finding new creative ways to trick internet users, be it in the form of lottery scams, downloading scam apps for smartphones or fake gambling websites. This paper presents a large-scale study on one particular delivery method of online scam: pop-up scam on typosquatting domains. Typosquatting describes the concept of registering domains which are very similar to existing ones while deliberately containing common typing errors; these domains are then used to trick online users while under the belief of browsing the intended website. Pop-up scam uses JavaScript alert boxes to present a message which attracts the user's attention very effectively, as they are a blocking user interface element.

Our study among typosquatting domains derived from the Alexa Top 1 Million list revealed on 8 255 distinct typosquatting URLs a total of 9 857 pop-up messages, out of which 8 828 were malicious. The vast majority of those distinct URLs (7 176) were targeted and displayed pop-up messages to one specific HTTP user agent only. Based on our scans, we present an in-depth analysis as well as a detailed classification of different targeting parameters (user agent and language) which triggered varying kinds of pop-up scams.

## KEYWORDS

phishing, typosquatting, scam, web security

## 1 INTRODUCTION

Pop-up ads have been an annoying phenomenon on the internet since the 1990s. This type of web advertisement puts the ad banner into a separate browser window instead of directly integrating it with the website [20]. The great popularity of pop-up ads among advertisement companies contrasted with the dissatisfaction of users because of the ads' highly intrusive nature. This eventually caused all major browser vendors to implement pop-up blockers in their software in the early 2000s [7]. Today, websites rarely make use of pop-ups and pop-up ads have disappeared almost completely from the web (as browsers would block them anyway).

However, similar concepts are now being used for online scams. Instead of displaying an ad or malicious content in a new browser window through the JavaScript method window.open,[1] a new trend in web-based scams can be observed: The JavaScript method alert[2] is used to show a short text message to the user.

Displaying the phishing message inside a JavaScript alert box has one important advantage for the attacker: An alert box steals the focus of the entire website. While normal advertisements can easily be ignored, alert boxes require the user to actively click a button to dismiss them. This obligatory interaction combined with the often short messages creates an effective entry point to further engage the user. This initial forced attention can then be exploited to lure

---

[1] https://developer.mozilla.org/en-US/docs/Web/API/Window/open (last accessed: 30 March 2019)
[2] https://developer.mozilla.org/en-US/docs/Web/API/Window/alert (last accessed: 30 March 2019)

the user to a dedicated website which serves the attacker's purpose, e. g. by asking for email addresses or credit card details. Attackers have also been observed repeatedly opening alert boxes, trying to pose as legitimate OS error messages and scaring the user into thinking that their device has been infected by malware [12].

These properties make alert boxes a very effective and widely abused vector for attackers. However, little attention has been paid to the described techniques by the research community. Based on the Alexa Top 1 Million websites [2], we created a list of websites with commonly misspelt names. This set consisted of 485 642 valid, registered domain names, which we scanned using automated browsers with five different user agents. In this paper, we present, to the best of our knowledge, the first comprehensive, large-scale study of the use of automatically displayed pop-up scams on websites and analyse how different user agents and languages are targeted by these campaigns.

In particular, the main contributions of this paper are:

- We present the first comprehensive scientific large-scale study of the utilisation of JavaScript pop-up messages for online scams on typosquatting URLs based on the Alexa Top 1 Million websites.
- We provide insight into the goals and purposes of the pop-up messages and the sites hosting them by manually defining and assigning categories based on the message content and the websites.
- Various distributions of the languages and the user agents across the different distinct messages, websites and categories are visualised and detailed, in order to explain the current state as well as trends in this particular delivery method for online scams.

The remainder of this paper is structured as follows: We discuss related work in section 2 and give a technical overview of the utilised framework in section 3. We present the results of our research in section 4 and evaluate the scan results in section 5, where we also present a large-scale analysis. Possible future work is detailed in section 6 and section 7 concludes the paper.

## 2 RELATED WORK

One important online scam category is phishing. It has been around for a long time as one of the most effective social engineering techniques and is a well-studied research area (see e. g. [1, 13, 14]). Due to the fact that the majority of today's users have only limited technical and security knowledge, the success rate of social engineering attacks is constantly high. Moreover, adversaries are becoming more and more creative in handcrafting their attacks to increase their success rate. While traditional means of delivery (i. e. via email [18]) are still widely used, many other delivery methods exist. Typosquatting [6] (also referred to as "URL hijacking") is a technique which is based on the concept of registering domain names with typing errors and similar mistakes made by users when entering a popular web address.

One of the first large-scale studies on typosquatting was conducted in 2003 by Edelman [5], who discovered more than 8 800 registered domains which were typographical variations of the most popular domain names at that time. His findings showed that most of those domain names were traced back to one individual, John Zuccarini, who used these typosquatted domains to redirect users to websites containing sexually explicit content. Furthermore, he was found to use particular tactics to trap the users from leaving these sites, such as blocking the browser's "Back" and "Close" functionalities.

Typosquatting attacks are based on the insertion, deletion or substitution of characters or the permutation of adjacent characters in popular domain names [9]. Holgers et al. [8] conducted an experiment in 2006 in which they measured the effect of visual similarities between letters in particular domain names. At that time, their results outlined that such homograph attacks were very rare and not severe in nature. However, the increasing use of internationalised domain names (IDNs) as well as the rising number of malicious IDN registrations over the last years show the increasing significance of this typosquatting technique [11, 21].

Numerous other squatting techniques such as *bitsquatting*, *combosquatting*, and *soundsquatting* were thoroughly researched in the past. Bitsquatting is the act of registering a domain name one bit different than an original domain, which might be accessed by users due to bit errors changing their memory content. Dinaburg [4] performed an experiment in which he registered 30 bitsquatted versions of popular domains and logged all HTTP requests. His findings outlined that there were 52 317 bitsquat requests from 12 949 unique IP addresses over the course of eight months. Nikiforakis et al. [17] conducted one of the first large-scale analyses of the bitsquatting phenomenon. Their results clearly showed that new bitsquatting domains are registered daily and are commonly used by the adversaries for generating profit through the use of ads, abuse of affiliate programs and, in some cases, distribution of malicious content.

Kintis et al. [9] conducted a study on combosquatting, which combines brand names with other keywords in the domain names. Their study showed that combosquatting domains are widely used to perform various types of attacks, including phishing, social engineering, affiliate abuse, trademark abuse and malware.

Furthermore, Nikiforakis et al. [15] presented a concept called soundsquatting which takes advantage of user confusion over homophones and near-homophones, i. e. words which sound similar or the same, but are spelled differently. To verify how much this soundsquatting technique is used in the wild, Nikiforakis et al. developed a tool to generate possible soundsquatted domains from a list of target domains. Using the Alexa Top 10,000 websites, they were able to generate 8 476 soundsquatted domains out of which 1 823 were already registered.

Additionally, Nikiforakis et al. [16] conducted a study in which they examined malicious JavaScript inclusions. Their findings included a vulnerability which occurs when a developer mistypes the address of a JavaScript library in their HTML pages. This would allow an attacker to easily register the typosquatted domain which could then compromise the website including a malicious JavaScript library.

Pop-up scam has not been researched in much detail yet. Miramirkhani et al. [12] performed a large-scale analysis of one particular type of pop-up scams, namely technical support scams. Their methodology included a check for JavaScript alert boxes. In Chou et al.'s work [3] the detection of traditional (JavaScript-less) pop-up ads through machine learning was proposed. The psychological aspects of fake pop-ups on internet users were analysed by Sharek et al. [22].

## 3 TECHNICAL OVERVIEW

To perform the large-scale scans required for this research, we employed a modified version of the MININGHUNTER [19] framework, which we initially developed to identify browser-based cryptocurrency mining campaigns. MININGHUNTER is based on Docker Swarm[3] and consists of automated browsers and a back end where the collected data is stored.

To scan websites at a large scale, a Chromium browser installed inside a Docker container is automated using the Chrome DevTools protocol.[4] It receives scanning requests via a Kue[5] job queue, automatically loads the website and records various details such as visited URLs. The accumulated data is then sent to a back end container through HTTPS and stored inside a MongoDB[6] database for later analysis. To scan a large number of websites within a reasonable time span, multiple scanning containers can be active at the same time.

For the purpose of testing we mimicked the most common behaviour of an adversary, namely, we made use of a technique popularly known as "typosquatting", as explained in section 2. In our experiment, we applied this technique to the Alexa Top 1 Million websites. To be able to cover the broad spectrum of the web address permutations, we used dnstwist,[7] a tool which generates possible typosquatting URLs for a particular URL. (At the time of performing this experiment, dnstwist only generated permutations of URLs; the tool's functionality has since been expanded significantly.) From the pool of thousands of possible address permutations, we selected only those which were actually registered as valid domains (in total, we were able to generate and verify 485 642 registered domain names).

For the purpose of this research, we developed two additional custom plugins for our framework. The first plugin, *UserAgentSpoofer*, sends a configurable, fake user agent to allow us to discern differences in behaviour which depend on this HTTP header. The plugin replaces the User-Agent request header in all requests sent to websites using the Network.setUserAgentOverride method of the Chrome DevTools protocol. The second plugin, *AlertRecorder*, stores URLs and messages of all JavaScript alert boxes encountered while loading and rendering a website. The data is acquired using the Page.javascriptDialogOpening API.

Websites are scanned until the Network.loadingFinished event is triggered by the Chrome DevTools protocol, plus an additional second in order to capture alerts that appear after the site has finished loading. The scan is also stopped in case the Network.loadingFinished event is not triggered 30 seconds after beginning to load the website.

Using these two plugins, we performed five full scans of our list of typosquatting domains based on the Alexa Top 1 Million websites. To be able to provide a wider variety of targets, each scan used a different user agent. We selected Chrome 69 (from 2018) and Firefox 46 (from 2016) to represent two popular, modern browsers running on Windows 10. We additionally included Internet Explorer 11 (from 2015) on Windows 7 to determine if any campaigns specifically target Microsoft's default browser for that OS. To cover the most commonly used mobile devices, we included Chrome 69 on Android 8.1 and Safari 12 on iOS 12 (both from 2018). Detailed information regarding all user agents selected for the scans can be found in Table 1.

## 4 RESULTS

Our scans (utilising different user agents as described in section 3) resulted in a total of 9 857 recorded alert boxes as well as 8 255 distinct URLs and 222 distinct messages. 8 828 of the recorded alert boxes can be considered malicious. An interesting aspect of our results is the targeting of specific user agents, which is further detailed in section 5: 7 176 websites displayed an alert box only to one particular user agent, whereas 1 079 websites showed messages to more than one user agent. Considering only distinct messages, we observed similar behaviour, although the difference is not as prominent – 126 distinct messages were only shown to one particular user agent, 96 to more than one.

---

[3] https://docs.docker.com/engine/swarm/key-concepts (last accessed: 30 March 2019)
[4] https://chromedevtools.github.io/devtools-protocol (last accessed: 30 March 2019)
[5] https://github.com/Automattic/kue (last accessed: 30 March 2019)
[6] https://www.mongodb.com/what-is-mongodb (last accessed: 30 March 2019)
[7] https://github.com/elceef/dnstwist (last accessed: 30 March 2019)

| label | user agent | operating system | browser |
| --- | --- | --- | --- |
| chrome | Mozilla/5.0 (Windows NT 10.0; Win64; x64) AppleWebKit/537.36 (KHTML, like Gecko) Chrome/69.0.3497.100 Safari/537.36 | Windows 10 | Chrome 69 |
| ie | Mozilla/5.0 (Windows NT 6.1; WOW64; Trident/7.0; rv:11.0) like Gecko | Windows 7 | Internet Explorer 11 |
| iossafari | Mozilla/5.0 (iPhone; CPU iPhone OS 12_0_1 like Mac OS X) AppleWebKit/605.1.15 (KHTML, like Gecko) Version/12.0 Mobile/15E148 Safari/604.1 | iOS 12 | Safari 12 |
| firefox | Mozilla/5.0 (Windows NT 10.0; WOW64; rv:46.0) Gecko/20100101 Firefox/46.0 | Windows 10 | Firefox 46 |
| androidchrome | Mozilla/5.0 (Linux; Android 8.1.0; TA-1053 Build/OPR1.170623.026) AppleWebKit/537.36 (KHTML, like Gecko) Chrome/69.0.3497.100 Mobile Safari/537.3 | Android 8.1 | Chrome 69 |

Table 1: The user agents used for the scans. "Label" is a unique identifier used throughout this paper when referring to the corresponding user agent. "Operating system" and "browser" refer to corresponding technology implied by the user agent.

## 5 EVALUATION

Using the categories described in subsection 5.1 as well as the user agents shown in Table 1, we determined specific characteristics of the recorded alert box messages with respect to these features.

### 5.1 Categories

In order to determine which websites try to achieve similar goals by displaying a message inside an alert box as well as to enable clearer visual representations of the distribution of message types across different user agents, we selected a number of categories from our findings and assigned one to each message.

Most messages in the FRAUD category declare that the user will receive some free credit to be used for gambling on the according website if they register and enter their debit card code, credit card information or similar data.

Messages contained in the LOTTERY category either claim the visitor has already won a lottery or that they have a particularly high chance of winning one. Such websites often either require the user to play a "game", such as spinning a wheel of fortune, or to answer questions regarding the prize (e. g. a smartphone). After completing such tasks, the websites reveal that the prize is actually a "special offer" and ask the visitor to provide their credit card information. Most of the messages in this category are in German; we explain this circumstance in subsection 5.4.

All messages in the category APK are in Chinese and most of them urge the user to download a dedicated application for displaying adult content. Unlike alert boxes in the category MOBILE CLIENT, they do not redirect to app store websites, but instead offer a direct download of an Android APK file or redirect to an iOS `itms-services` URL. Several samples were analysed using VirusTotal[8] and were identified as potentially unwanted programs (such as adware and spyware) as well as Trojans.

Based on the characteristics of the alert box message content as well as manual inspection of selected samples for each distinct message, we consider messages inside the categories FRAUD, LOTTERY, and APK to be malicious (e. g. phishing). Besides these malicious categories, we further defined various non-malicious categories; they were differentiated by content and message purpose in order to gain additional insight into the reasons for showing alert boxes in general.

The category ERRORS contains several types of error messages, e. g. indicating invalid access tokens or JavaScript errors as well as website maintenance and discontinuation notices.

---

[8] https://www.virustotal.com (last accessed: 30 March 2019)

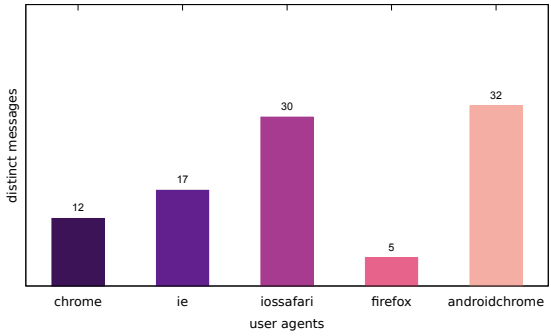

Figure 1: Number of distinct messages displayed to one particular user agent

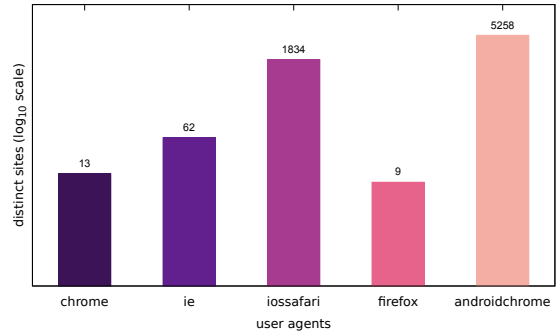

Figure 2: Number of websites targeting one particular user agent

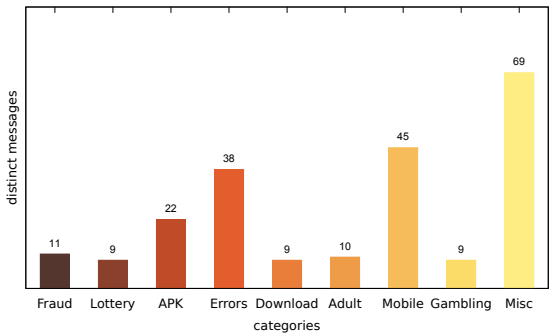

Figure 3: Number of distinct messages per category

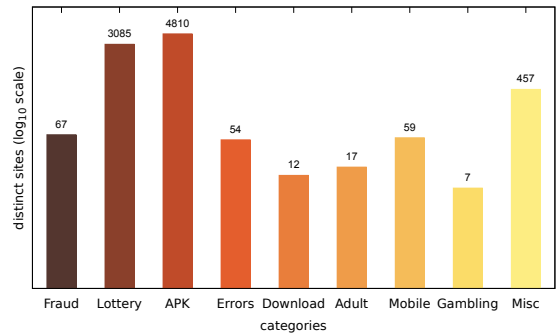

Figure 4: Number of distinct sites per category

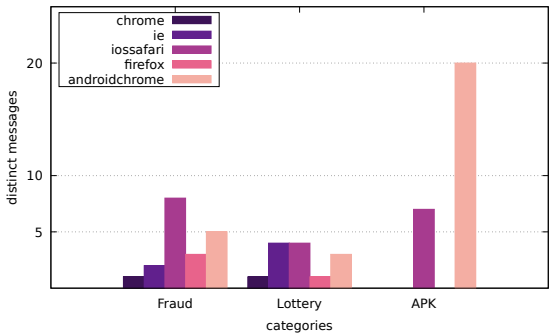

Figure 5: Distribution of distinct messages over different user agents in malicious categories

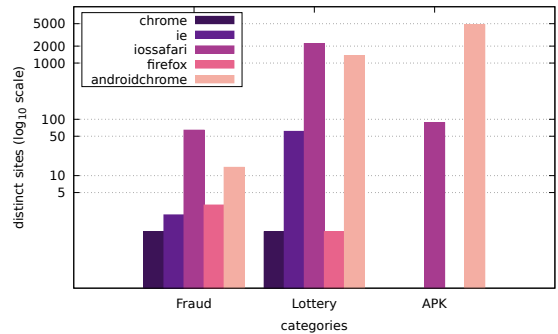

Figure 6: Distribution of distinct sites over different user agents in malicious categories

Messages categorised as DOWNLOAD urge the user to install or update either Java or Adobe Flash Player and redirect the user to the corresponding download area. Manual inspection of the websites included in our scan which displayed these messages showed that the alert boxes do actually redirect to the legitimate websites of the software manufacturers.

ADULT messages inform the user about adult content on the visited website, ask the user to confirm that they are of legal age and present the website's terms and conditions.

Messages of the category MOBILE SITE ask the visitor whether they want to display the dedicated mobile version of the website.

Mobile Client messages inform the user about the website's smartphone app and redirect the user to the according app store website.

The categories Mobile Site and Mobile Client are combined into the category Mobile in diagrams throughout this paper.

Messages of the Gambling category are related to gambling websites. All messages are in Chinese, and most websites hosting these alert boxes provide the latest results of the Hong Kong Jockey Club's Mark Six lottery[9] as well as other gambling information. They either require the user to register on a different website or present a special offer along with an ID or contact number for instant messengers, which are in widespread use in the People's Republic of China. Since the websites did not directly request credit card information or deceive the visitors in other ways, and since we could not easily investigate the associated instant messenger accounts, we chose to separate these messages into the category Gambling instead of including them in the more explicitly malicious category Lottery.

Misc categorises alert box contents which do not fit into any other category and include short cookie policy statements, welcome messages and password prompts as well as various other kinds of miscellaneous messages.

## 5.2 Analysis

Our results show that a significant portion of the scanned websites target visitors with mobile web browser user agents. As Figure 1 illustrates, there are few distinct messages displayed only to a specific user agent. While the difference in the number of messages only shown to one specific user agent is not significant, there is a large disparity between the number of websites focussing on desktop web browsers and those targeting mobile web browsers in general.

Figure 2 illustrates the number of websites which displayed an alert box only to one particular user agent. While some alert boxes are legitimately directed at specific user groups, the number of alert boxes shown only to desktop browsers is almost negligible. We present some possible explanations for the specific targeting of mobile users in subsection 5.1 as well as subsection 5.4.

As we found many websites displaying alert boxes only to users with mobile web browser user agents and a relatively low diversity of messages (many of which are presented to one particular user agent only), our findings indicate that there are relatively few operators deploying their resources on a large number of different websites at the same time.

---

[9] https://bet.hkjc.com/marksix/ (last accessed: 30 March 2019)

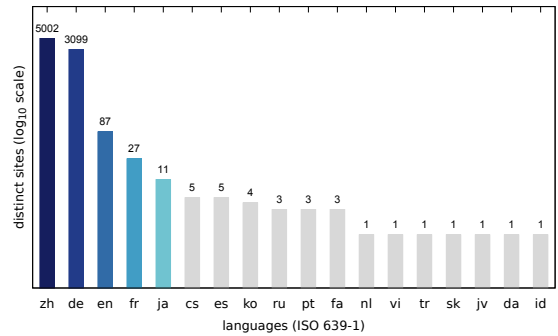

Figure 7: Number of distinct sites displaying messages in a specific language

## 5.3 Category Analysis

To make further analysis of the message content possible as well as to simplify the identification of specific phishing campaigns, we translated every message into English utilising Google Translate. This allowed us to classify the messages into the content categories described in subsection 5.1 regardless of the original language.

Figure 3 shows the number of distinct messages in each category; as the figure shows, the greatest diversity in message content occurs in the categories Misc, Mobile and Errors.

While the majority of distinct messages belongs to legitimate non-malicious categories, most of the recorded alert boxes actually do fall into malicious categories. Figure 4 depicts the number of sites in each category. The vast majority of alert boxes belong to the categories APK and Lottery, either trying to trick users into downloading and installing smartphone apps outside of the controlled environment of their OS's application store or trying to trick users into providing their credit card information by promising them some kind of lottery prize. The large number of distinct messages and websites in the Misc category is attributable to the scattered characteristics of the messages in this category and is therefore not as significant.

Finally, we want to discuss the joint distribution of user agents and message content. Figure 5 displays the number of distinct messages shown only to a particular user agent for each of the malicious categories (Fraud, Lottery, and APK), whereas Figure 6 shows the corresponding graph for distinct websites. Most malicious alert boxes were encountered while utilising a mobile web browser user agent, while only a small fraction of the websites showed alert boxes on desktop browsers, at all.

Our findings indicate that the majority of websites in the categories Lottery and Fraud targeted the user agents iossafari and androidchrome (in that order), while the websites in the category APK exclusively targeted user agents of mobile web browsers (for obvious reasons), with a strong focus on the user agent androidchrome. The strong preference for targeting Android can be explained by the relatively simpler process of installing apps from outside the app store on Android phones [10], whereas installing apps from unknown sources on iOS requires a more complicated procedure (such as first installing an enterprise certificate on the smartphone [23]).

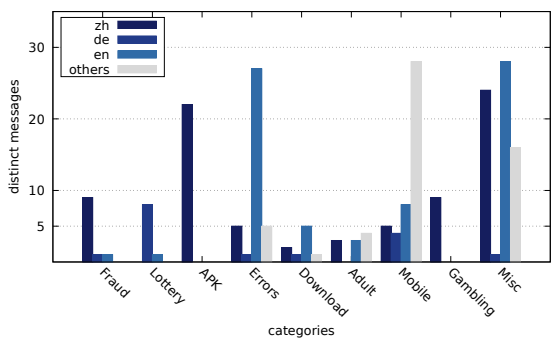

Figure 8: Distribution of distinct messages over different languages by category

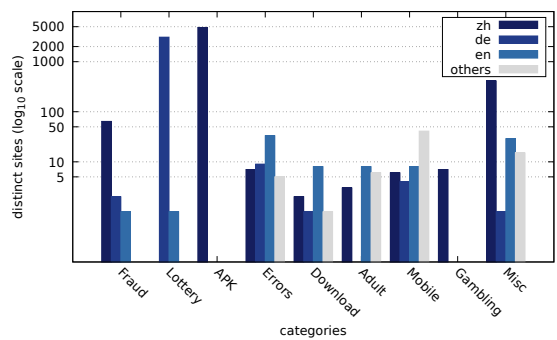

Figure 9: Distribution of distinct sites over different languages by category

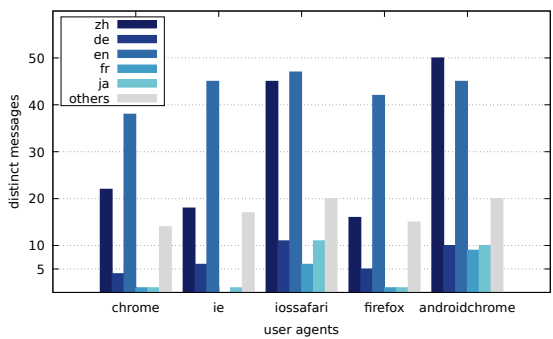

Figure 10: Distribution of distinct messages over different languages by user agent

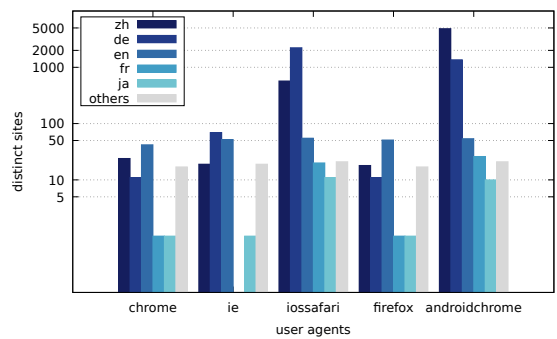

Figure 11: Distribution of distinct sites over different languages by user agent

## 5.4 Language Distribution

Since we discovered a number of different trends for targeting specific user groups, we additionally analysed the language distribution of the collected messages. As shown in Figure 7, the vast majority of websites displayed messages written in either Chinese (zh) or German (de). The main reason for the large number of Chinese messages is evident in Figure 8 and Figure 9, which show the distribution of languages over messages as well as websites in the different categories: The category APK (which contains the largest number of websites, cf. Figure 4) consists solely of messages in Chinese, and several other categories have a relatively large fraction of messages in Chinese, as well.

As Figure 8 shows, there are not as many distinct messages in German across the different categories. However, Lottery (which contains the second largest number of websites, cf. Figure 4) consists primarily of messages in German (cf. Figure 9). Based on "untranslated" parts (e. g. the currency units) as well as hints of localisation such as a town name and the name of a large German consumer electronics retailer (especially in the category Lottery), we suspect that the large number of websites displaying messages in German is likely a result of localisation attempts based on the Austrian IP address used for our scan. As mentioned in section 6, further scans utilising IP addresses of other countries and varying language settings are necessary in order to validate this hypothesis.

The distribution of the languages across the different user agents is depicted in Figure 10 as well as Figure 11. While the number of distinct English messages is large, only a relatively small fraction of the websites displays these messages. In summary, our results show that most alert boxes are displayed to visitors utilising web browsers with mobile user agents as well as that the majority of those alert box messages are in Chinese or German.

## 6 FUTURE WORK

Our work could be expanded upon in several directions. For one, the detection of the targeting scope could be extended to include more HTTP user agents.

Additionally, exploring the language- and location-specific targeting further by repeating the scans with varying language and location/IP settings (e. g. using VPN services) seems likely to bring about further insight into location-specific targeted phishing campaigns and effects on language-specific message content (cf. the preliminary findings in this direction in subsection 5.4).

The introduction of internationalised domain names (IDN) in 2010 introduced new attack vectors beyond simple typosquatting in the form of IDN homograph attacks, using homographs such as Greek omicron 'o' or Cyrillic es 'c'; we consider further research into the prevalence of IDN homograph attacks an interesting subject of study.

Finally, the manual process of assigning categories to distinct messages could be replaced by a fully automated classification process utilising machine learning algorithms. A review of the existing categories might be necessary, as well, and the categorisation might rely solely on the message content instead of including background knowledge. This new process could establish a periodical automatic analysis enabling the observance of developments and trends in pop-up scams.

## 7 CONCLUSION

Techniques similar to those used for displaying pop-up ads in the early days of the World Wide Web are now used by malicious websites to deliver online scam. JavaScript alert message boxes steal the focus of the website, show a short text message to the user and try to either lure or scare the user into taking specific actions or exposing their data. Unfortunately, little scientific attention has been paid so far to the techniques utilised by scam websites to gain the attention of users and to retrieve data such as credit card information.

We performed large-scale scans of typosquatting URLs based on the Alexa Top 1 Million websites via automated Chromium browsers utilising a modified version of the MiningHunter [19] framework. The scans with five different user agents resulted in a total of 9 857 recorded alert boxes, out of which 8 828 can be considered malicious.

Our in-depth analysis presented characteristics of web-based scam campaigns and outlined target groups and goals of the various attacks. It showed that a majority of websites displayed a pop-up box to one specific HTTP user agent only, and that most of them focused on mobile web browsers.

Different message categories were defined based on the message content and the websites displaying an alert box containing the message. The largest categories are Lottery and APK, which are trying to trick the user by making them believe they have won a lottery or to directly download and install a potentially malicious application, respectively.

Another aspect of our analysis was the distribution of different languages. We found that most websites were displaying alert box messages in Chinese, followed by German (which was most likely the result of IP-based location targeting). Chinese messages often fell into the category APK and targeted a mobile web browser user agent.

## ACKNOWLEDGEMENTS

This research was funded by the Austrian Research Promotion Agency (FFG) BRIDGE project 853264 "Privacy and Security in Online Advertisement (PriSAd)" as well as the Josef Ressel Center TARGET. The financial support by the Austrian Research Promotion Agency, the Federal Ministry for Digital and Economic Affairs and the Christian Doppler Research Association is gratefully acknowledged.